# Is Teleportation a (quantum) mystery?

Since its discovery quantum teleportation has often been seen as a manifestation, indeed the epitome, of the very paradoxical and mysterious nature of quantum theory itself. It is commonly regarded as genuinely quantum and essentially paradoxical. Although a common approach to teleportation amongst physicists nowadays is a somewhat operational one, some researchers are making an effort to deflate the above views.

On the one hand, it was recently argued that the paradox of information transfer taking place in teleportation is dissolved (Timpson, 2006) by appealing the very notion of information. On the other hand, it was demonstrated that some classical versions of teleportation retain its important features, which hitherto were considered genuinely quantum (Cohen, 2003; Collins&Popescu, 2002; Hardy, 1999; Mor, 2006; Spekkens, 2007).

I will present a special version of a quantum teleportation protocol which is in a sense split into classical and quantum steps. This description provides us with a unified picture of teleportation in both domains. It will be explicitly shown how classical teleportation is embedded in the quantum protocol. Moreover, the classical step can be successfully accomplished even if the state shared by the parties is completely disentangled [this is consistent with the result obtained in (Wang, 2005)]. Yet, all the (apparent) paradoxical features usually associated with quantum teleportation are clearly present in this step. In particular, this demonstrates that entanglement cannot be ultimately responsible and not necessary for the (paradoxical?) information transfer.

Thus, even if one considers teleportation as mysterious, all its mysteries are shifted from quantum domain into purely classical one.

## 1. Introduction

The original quantum teleportation proposal (Bennett, Brassard, Crepeau, Jozsa, Peres, and Wootters, 1993) had caused a good deal of excitement not only amongst quantum physics community, but amongst general public. The latter fact may seem somewhat surprising. Indeed, the main thrust of Star Track fans, i.e. disembodied "instantaneous transportation of persons, etc., across space" (Vaidman, 1998), is hardly achieved by quantum teleportation (QT). QT does not transport stuff at all. It requires presence of physical substance in both locations, and only the physical system's state of affairs is transported from one location to another. It is not instantaneous either. One can easily imagine a classical counterpart of QT that achieves the same goal for classical systems (Collins&Popescu, 2002). Therefore, QT should surprise a layman not more than the possibility to conduct a telephone conversation across the Atlantic: indeed, the voice of my colleague in New York does not cross the ocean, but the physical characteristics of sound waves generated by my telephone handset in England are (nearly) identical to those generated by her vocal cords in New York. Moreover, it can be arranged, in principle, (with the help of two computers with special pre-shared random data) that the electric signal that is actually travelling across the Atlantic during the conversation does not carry any information about those characteristics at all. The latter consideration highlights one of the apparent paradoxes that is normally associated with QT, but as we show here is shared by its classical analogue – if the electric signal that crosses the Atlantic does not carry information about my colleague's voice, then how the information needed to reproduce her voice gets from New York to Cambridge after all?

The aim of this article is not to show how paradoxes like the one presented above should be solved or dissolved, but rather to demonstrate that they do not pose any essentially quantum challenge and should be treated inside classical domain.

## 2. The puzzles of Quantum Teleportation

The simplest QT protocol (Bennett *et all*, 1993) can be presented in the following abstract form. Alice wants to teleport particle *a*, the state of which is unknown to her, to Bob. Most generally its state can be characterized by two real parameters, say $\theta$ and $\varphi$, which specify the position of the state-vector on the Bloch-sphere:

$$|\psi\rangle_a = \cos\theta |0\rangle_a + e^{i\phi} \sin\theta |1\rangle_a. \quad (0.1)$$

Throughout the article we will use the standard in quantum information theory *computational basis* notation, where $\{|0\rangle, |1\rangle\}$ denotes the orthonormal basis of a 2-dimentional Hilbert space corresponding to eigenvalues of $\sigma_z$ Pauli operator.

In addition, Alice and Bob share two particles *A* and *B* in a maximally entangled state

$$|\Phi^+\rangle_{AB} = \frac{1}{\sqrt{2}}\left(|0\rangle_A \otimes |0\rangle_B + |1\rangle_A \otimes |1\rangle_B\right). \quad (0.2)$$

Thus, Bob holds one particle, *B*, in his laboratory, while Alice posses two particle *a* and *A* in hers. We assume that Alice can perform any join transformation or measurement on these two particles. To facilitate teleportation Alice performs a projection measurement in the entangled basis, that spans the 2x2-dimentional Hilbert-space of *a* and *A*, known as the Bell basis:

$$|\Phi^\pm\rangle_{AB} = \frac{1}{\sqrt{2}}\left(|0\rangle_A \otimes |0\rangle_B \pm |1\rangle_A \otimes |1\rangle_B\right),$$
$$|\Psi^\pm\rangle_{AB} = \frac{1}{\sqrt{2}}\left(|0\rangle_A \otimes |1\rangle_B \pm |1\rangle_A \otimes |0\rangle_B\right), \quad (0.3)$$

The advantage in performing this measurement becomes evident as soon as one expands the original state of all three particles in term of the Bell-basis

$$|\psi\rangle_a \otimes |\Phi^+\rangle_{AB} = \frac{1}{2}|\Phi^+\rangle_{aA} \otimes \left(\cos\theta|0\rangle_B + e^{i\phi}\sin\theta|1\rangle_B\right)$$
$$+ \frac{1}{2}|\Phi^-\rangle_{aA} \otimes \left(\cos\theta|0\rangle_B - e^{i\phi}\sin\theta|1\rangle_B\right)$$
$$+ \frac{1}{2}|\Psi^+\rangle_{aA} \otimes \left(\cos\theta|1\rangle_B + e^{i\phi}\sin\theta|0\rangle_B\right)$$
$$+ \frac{1}{2}|\Psi^-\rangle_{aA} \otimes \left(\cos\theta|1\rangle_B - e^{i\phi}\sin\theta|0\rangle_B\right).$$

(0.4)

Depending on the result of the Alice's measurement different state of Bob's particle is *realized*[1]. The result $\Phi^+$ implies that the target state is realized. However, any of the three other possible states is related to the target state via a standard rotation, $\sigma_x, \sigma_y$ or $\sigma_z$, which can be implemented by Bob if he knows the result of the Alice's measurement. Thus, as there are only four possibilities it is sufficient if Alice sends to Bob 2 bits of information.

The paradox usually raised in relation to the above procedure is based on the following premise. The original quantum state of the particle *a* is parameterised by two real parameters, i.e. an infinite amount of information is needed to describe the state. Since at the end of the protocol the same state of the particle *B* is realised, the information about this state should have passed somehow from *a* to *B*. The way one tackles this problem depends on the one's interpretation of the notion of information and interpretation of a quantum state.

One view is to deny the very premise (Timpson, 2006). Timpson argues that information is not a physical substance, but rather is an abstract noun and is not expected to follow a continuous spatio-temporal path. Thus, according to Timpson, there is no need to show *how* information gets from Alice to Bob.

The opposite, more traditional, view accepts the premise and the efforts are made to find continuous spatio-temporal paths taken by the information about $|\psi\rangle$. It is useful to distinguish two aspects of the information transfer paradox as it highlights the difference between the views of those who hold different interpretations of a quantum state. The first aspect of the paradox is related to the fact that the results of Bell-measurement are random, and therefore the 2 bits sent from Alice to Bob are random as well, they are completely unbiased and do not contain any information whatsoever about the original state. Thus, the first question is *how does the information get from Alice to Bob*? Those who assign ontological meaning to a single specimen of quantum state and those who assign such a meaning only to an ensemble of states are concerned with this question. The second aspect arises due to observation that even if the 2 bits did carry some information about the original state, they still could not carry the whole (infinite) amount of information needed to describe two real parameters. Thus, the second question is *how is so much information transported*? The latter question does not worry those who assign an ontological meaning to a quantum state only at ensemble level. Teleporting many copies of a state would mean sending many bits of classical information anyway.

In the next Section I will discuss the simplest classical analogy of QT.

---

[1] I avoid using the term *collapsed* to maintain my analysis as interpretation-free.

## 3. One-time pad as a "classical teleportation"

One-time pad (OTP) is a primitive (classical) cryptographic protocol, where two parties use pre-shared secret key (two strings of random correlated bits – secret classical correlations) to securely transmit a string of data bits. OTP was first analysed as a classical analogy of QT by Collins and Popescu (2002) and discussed by Mor (2006). The protocol works as follows. Let us denote the data bit, which is initially held by Alice by $a$[2] and the two correlated bits by $A$ and $B$. Alice performs the logical XOR gate on $a$ and $A$ (sum modulo 2), which effectively determines parity of the two bits. The result, $a \oplus A$, is completely random unbiased bit, which can be safely communicated to Bob via a public channel. To recover the initial bit on his side Bob performs XOR on the communicated bit and his bit $B$.

|       | $a$ | $A$ | $B$ | $a \oplus A$ | $(a \oplus A) \oplus B$ |
|-------|-----|-----|-----|--------------|-------------------------|
| $p$   | 0   | 0   | 0   | 0            | 0                       |
|       |     | 1   | 1   | 1            |                         |
| $1-p$ | 1   | 0   | 0   | 1            | 1                       |
|       |     | 1   | 1   | 0            |                         |

Table 1: The truth-table for OTP, where $a$ has a probability distribution {p, 1-p}.

The key point is that this protocol shares the most important features with QT.

First, the classical bit $a \oplus A$, which Alice communicates to Bob is completely unbiased, i.e. its value is statistically independent on the value of $a$ (and on its probability distribution *{p,1-p}*). This fact raises a similar question of how does the information about the value of *a* gets from Alice to Bob.

Second, although the probability parameter *p* does not characterise the state of *a* in the same sense as *θ* and *φ* characterise the state of particle *a* in the previous section, the information about *p* can be interpreted as being transported to *B*. Indeed, in the standard information-theoretical picture the bit *a* is generated by some source. The probability distribution *{p,1-p}* is a physical characteristic of that source. OTP allows Bob to *sample* from the source directly. In other words, the source becomes accessible to Bob directly[3]. This claim will be substantiated in the next section, where it will be shown how a classical teleportation step (essentially equivalent to OTP) emerges when QT protocol is modified in a special way.

There is a common belief that the *no-cloning theorem*, which prohibits making perfect copies of an unknown quantum state, is the very aspect that makes QT essentially different from OTP. I find this idea erroneous. Despite the fact, that the procedure described above does not require the original copy of *a* to be preserved, Alice may implement (irreversible) one-output XOR gate without keeping a copy, if she wishes to do so. The result of her parity measurement will not depend on whether she did or did not keep a copy. It is a crucial conclusion, that the possibility to keep a copy *does not facilitate* OTP. The possibility of cloning is, certainly, a feature of OTP, but it is not an essential feature as far as teleportation is concerned. Whether no-cloning crucial or not, classical analogues of QT that preserve the no-cloning feature are possible. Hardy (1999) and Spekkens (2004) have demonstrated teleportation in classical toy theories with no-cloning.

---

[2] Here variables and their values are denoted by a same symbol.

[3] The similar view was advocated by Cohen (2003) and Mor (2006).

### 4. Two-step teleportation protocol

In this section we demonstrate how the standard teleportation protocol can be decomposed into equivalent two-step protocol, where the first step will be associated with teleportation of a classical part of a state and the second step with its coherent phase. We replace the four outcome Bell measurement, i.e. a projective measurement in the entangled basis, the Bell-basis, by unitary operation CNOT acting on the two qubits followed by two separate measurements on these qubits, $\sigma_z^A$ and $\sigma_x^a$.

It is obvious that Alice can treat the two measurements (of *a* and *A*) as a single measurement in a product basis, in which case she effectively implements the Bell measurement, but in a different way. The interesting case, however, is when she treats these measurements separately.

*Step 1: Alice applies CNOT and measures particle, A, in the z-basis.* The CNOT transformation acts as follows

$$U_{CNOT}^{a \multimap A}|\psi\rangle_a \otimes |\Phi^+\rangle_{AB} = \cos\theta|0\rangle_a \otimes |\Phi^+\rangle_{AB} + e^{i\phi}\sin\theta|1\rangle_a \otimes |\Psi^+\rangle_{AB}$$
$$= \frac{1}{\sqrt{2}}|0\rangle_A \otimes \left(\cos\theta|0\rangle_a \otimes |0\rangle_B + e^{i\phi}\sin\theta|1\rangle_a \otimes |1\rangle_B\right) \quad (0.5)$$
$$+ \frac{1}{\sqrt{2}}|1\rangle_A \otimes \left(\cos\theta|0\rangle_a \otimes |1\rangle_B + e^{i\phi}\sin\theta|1\rangle_a \otimes |0\rangle_B\right),$$

where $U_{CNOT}^{a \multimap A} = |0\rangle_a\langle 0| \otimes I_A + |1\rangle_a\langle 1| \otimes \sigma_x^A$.

The measurement of particle *A* in the computational basis leaves particles *a* and *B* in an entangled state. As a result, the reduced density matrix of *B* will be $\rho_B = (\cos\theta)^2|0\rangle_B\langle 0| + (\sin\theta)^2|1\rangle_B\langle 1|$ or $\sigma_x\rho_B$ depending on the outcome of the measurement. Alice informs Bob about this outcome by communicating one bit of information to him, which allows Bob to apply a correction in the latter case. From Bob's point of view, the original state was partially teleported to him already in the following sense. The classical part, i.e. the part represented by the projection of the pure state onto z-axis, was teleported. In other words, not the original state, but its *decohered version* was teleported. At this stage, one real parameter, *θ*, describes the state of Bob's particle (see Fig. 1).

*Step 2: Alice measures her original particle, a, in the x-basis.* Let us rewrite the state of *a* and *B* in terms of the basis states of *a* corresponding to the Alice's measurement:

$$|\psi\rangle_{aB} = \cos\theta |0\rangle_a \otimes |0\rangle_B + e^{i\phi}\sin\theta |1\rangle_a \otimes |1\rangle_B = \frac{1}{\sqrt{2}}|\rightarrow_x\rangle_a \otimes \left(\cos\theta|0\rangle_B + e^{i\phi}\sin\theta|1\rangle_B\right)$$
$$+ \frac{1}{\sqrt{2}}|\leftarrow_x\rangle_a \otimes \left(\cos\theta|0\rangle_B - e^{i\phi}\sin\theta|1\rangle_B\right),$$

where $|\rightleftarrows_x\rangle \equiv (|0\rangle \pm |1\rangle)/\sqrt{2}$. Alice communicates her result to Bob who corrects his state, if necessary, by applying $\sigma_z$ rotation, if necessary. This completes the full teleportation protocol.

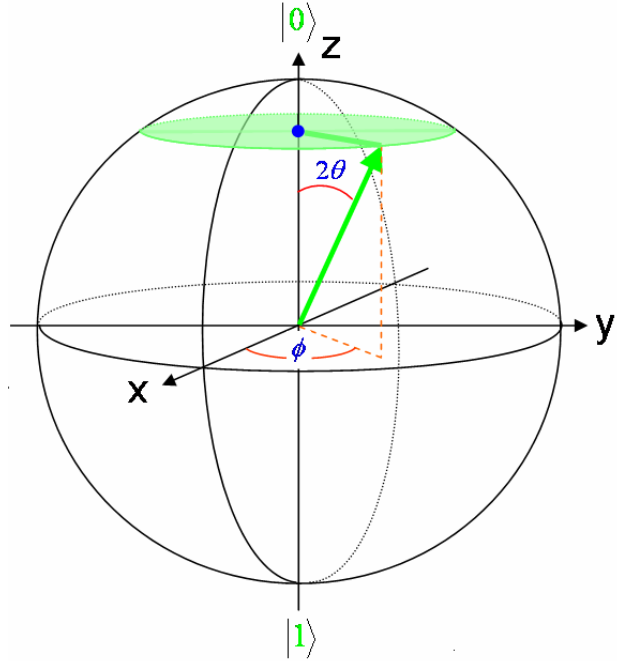

Figure 1: The two-step evolution of a single-particle state. Initially it is pure, $|\psi\rangle_a$ (the tip of the arrow). As a result of the Step 1 it decoheres, i.e. projects onto the z-axis (the bold point). One parameter, namely $\theta$, suffices to describe its position on z-axis. As a result of the Step 2 the second parameter, $\varphi$, is transferred and full coherence is restored – the state returns to the surface of the sphere.

Technically speaking the above procedure is a trivial manipulation by a 3-particle system via LOCC, which manifest itself in swapping entanglement from particles A and B to particles $a$, $B$ at Step 1, followed by disentanglement of all particles completely resulting in teleportation of the original state of $a$ to $B$ at Step 2 (see Fig. 2). Nevertheless, the analysis of the apparent "flow" of information about the parameters $\theta$, $\varphi$ can yield important insight.

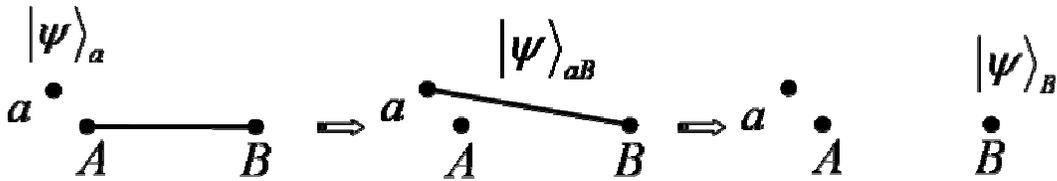

Figure 2: The two-step teleportation scheme, where the initial pure state $|\psi\rangle$ is transferred from the Hilbert space of $a$ to the product Hilbert space of $a$ and $B$ at Step 1 before finally being confined to the Hilbert space of $B$ only.

Consider Step 1 again. As its outcome we obtain the following transformation

$$\left(\cos\theta|0\rangle_a + e^{i\phi}\sin\theta|1\rangle_a\right)\otimes|\Phi^+\rangle_{AB} \to |\chi\rangle_A \otimes\left(\cos\theta|0\rangle_a\otimes|0\rangle_B + e^{i\phi}\sin\theta|1\rangle_a\otimes|1\rangle_B\right), \quad (0.6)$$

where $|\chi\rangle_A$ is either $|0\rangle$ or $|1\rangle$. Let us assume, for a moment, that we accept the ontological reality of a single specimen of a quantum state. Initially, the whole information about the state $|\psi\rangle$, i.e. about $\theta$ and $\varphi$, is located at Alice's site [LHS of Eq. (0.6)]. Our interpretation should be consistently applied to the joint state of *a* and *B* in the RHS of Eq. (0.6), however it is harder to say where, if at all, this information is localised. Because $\theta$ and $\varphi$ parameterise a non-local state, the information about them is in a sense delocalised. Nevertheless, $\theta$ is accessible also locally, though only at an ensemble level as the reduced density matrix on Bob's (and similarly on Alice's side) is

$$\rho_B = (\cos\theta)^2 |0\rangle_B\langle 0| + (\sin\theta)^2 |1\rangle_B\langle 1|. \quad (0.7)$$

Notice, that $\varphi$ does not appear in Eq. (0.7), i.e. unlike $\theta$ the parameter $\varphi$ is completely delocalised. In other words, $\varphi$ became a non-local phase, and it is no longer locally accessible to either Alice or Bob neither at single-copy or ensemble level.

As we have already stated before the one's attitude towards interpretation of the above procedure depends on the one's interpretation of a quantum state. The information about $\theta$ might be accessible to Bob (and Alice) only at ensemble level, but it will be rather inconsistent to assign an ontological meaning to a single specimen of the pure entangled state of *a* and *B*, but not to a single specimen of the reduced state $\rho_B$. This states is a mixed state, but the mixture is *improper*, and therefore it is hard to deny that even a single specimen of $\rho_B$ is an ontic state in it's own right (if it's pure extension is). In this very sense the information about $\theta$, which parameterises $\rho_B$, is transported to *B*.

Step 1 can be interpreted as a *classical* step of the QT protocol for two reasons. First, it teleports only completely decohered version of the original state, which is effectively classical. In other words it 'filters' the state's classical part. Second, from Bob's local point of view Step 1 can be accomplished with no entanglement at all! A completely disentangled state, representing effectively classical correlation[4] will suffice:

$$\left(\cos\theta|0\rangle_a + e^{i\phi}\sin\theta|1\rangle_a\right)\otimes\left(\frac{1}{2}|00\rangle\langle 00|_{AB} + \frac{1}{2}|11\rangle\langle 11|_{AB}\right)$$
$$\to |\chi\rangle_A \otimes\left([\cos\theta]^2|00\rangle\langle 00|_{aB} + [\sin\theta]^2|11\rangle\langle 11|_{aB}\right),$$

as the reduced density matrices are still the same as in Eq. (0.7). Here the interpretation of $\theta$ becomes even more profound.

---
[4] The correlated state in the equation below is purely classical. It should be obvious in this particular case, however see [Groisman, Kenigsberg and Mor, (2007)] for justification.

On the Step 2, all three particles become disentangled and both parameters $\theta$ and $\varphi$ are localized in B:

$$|\chi\rangle_A \otimes \left(\cos\theta |0\rangle_a \otimes |0\rangle_B + e^{i\phi}\sin\theta |1\rangle_a \otimes |1\rangle_B\right) \to |\chi\rangle_A \otimes |\xi\rangle_a \otimes \left(\cos\theta |0\rangle_B + e^{i\phi}\sin\theta |1\rangle_B\right) \quad (0.8)$$

where $|\xi\rangle_a = |\rightleftarrows_x\rangle_a$.

From the above discussion it might appear that $\theta$ and $\varphi$ parameterise the *classical* and the *quantum* part of $|\psi\rangle$ respectively. However, such a conclusion should be avoided. Classicality of $\theta$ and quantumness of $\varphi$ have a relative meaning, which depends on the choice of the original computational basis, i.e. basis in which the state decoheres. The whole procedure could have been performed in a different basis, in which case $|\psi\rangle$ would have been parameterised in a different way.

## 5. Classical delocalization of information

Quite remarkably we do not tend to associate the question of the spatio-temporal discontinuity of the information-path with the classical information processing. Yet, it manifests itself even in simplest cases. For example, it is well known that a value of any binary variable $d$ can be encoded in the correlation between two completely random variables. How is it possible? Consider two binary variables $x$ and $y$, which are correlated in the following way $x \oplus y = 0$ (i.e. they are either both 0 or both 1 with equal probability). This type of correlation is often called *secret random classical correlation* or shared randomness (similar to one we used in Sec. 3 for OTP). Everything we need to do in order to encode $d$ in the correlation between two variables is to apply a XOR gate on $d$ and $x$ and update $x$ according to the rule $\tilde{x} = d \oplus x$. The updated value $\tilde{x}$ is random, which is evident from the truth-table

|      | $d$ | $x$ | $y$ | $\tilde{x} = d \oplus x$ | $\tilde{x} \oplus y$ |
|------|-----|-----|-----|--------------------------|----------------------|
| $p$  | 0   | 0   | 0   | 0                        | 0                    |
|      |     | 1   | 1   | 1                        |                      |
| $1-p$| 1   | 0   | 0   | 1                        | 1                    |
|      |     | 1   | 1   | 0                        |                      |

Indeed, the probability $P(\tilde{x} = 0)$ is

$$P(\tilde{x}=0) = P(\tilde{x}=0/d=x=0)P(d=x=0) + P(\tilde{x}=0/d=0, x=1)P(d=0, x=1)$$
$$+ P(\tilde{x}=0/d=1, x=0)P(d=1, x=0) + P(\tilde{x}=0/d=x=1)P(d=x=1)$$
$$= 1 \times \frac{p}{2} + 0 \times \frac{p}{2} + 0 \times \frac{(1-p)}{2} + 1 \times \frac{(1-p)}{2} = \frac{1}{2}.$$

Thus, the original value *d* is encoded in the parity of $\tilde{x}$ and *y*. Yet, individually both variables are still completely random. Moreover, we did not actually have to alter the value of *y*! This trivial observation is remarkable, because *d* and *x* could be located with Alice, while *y* could be with Bob who is light-years away. The space-like separation would not prevent the information about the value of *d* from *delocalizing instantaneously* via encoding into the parity of *x* and *y*. One does not need quantum non-locality (in fact, quantum at all) to make *d* locally inaccessible. Despite so many discussion of teleportation, no one was ever puzzled by *this*. The procedure can be reversed and *d* can be localized again using OTP, i.e. utilizing shared random correlations and sending one random bit.

The above example is compelling evidence that even in purely classical data processing information path can be spatio-temporally discontinued. This phenomenon is of essentially classical nature and does not present any quantum puzzle.

## 6. Summary and conclusion

We have shown that despite the common belief QT does not posses any genuinely quantum puzzles. This does not imply, however, that QT is a classical procedure, which it is definitely not! It does imply that all the problems that are traditionally associated with QT can be identified with a classical domain.

## Funding
This work was funded by the U.K. Engineering and Physical Sciences Research Council, Grant No. EP/C528042/1, and supported in part by the European Union through FP6-FET Integrated Projects SCALA (CT-015714) and QAP (CT-015848).

## Acknowledgements
I am grateful to Jeremy Butterfield for his valuable remarks, encouraging discussions and support. I would like to thank Rob Spekkens and Matthew Leifer for helpful comments and suggestions.

Berry Groisman
*Centre for Quantum Computation*
*Department of Applied Mathematics and Theoretical Physics,*
*University of Cambridge,*
*Cambridge CB3 0WA, UK*
*b.groisman@damtp.cam.ac.uk*